\documentclass[fleqn,usenatbib]{mnras}

\usepackage{newtxtext,newtxmath}

\usepackage[T1]{fontenc}

\DeclareRobustCommand{\VAN}[3]{#2}
\let\VANthebibliography\thebibliography
\def\thebibliography{\DeclareRobustCommand{\VAN}[3]{##3}\VANthebibliography}

\usepackage{graphicx}	
\usepackage{amsmath}	
\usepackage{amssymb}	

\usepackage{graphicx}	
\usepackage{amsmath}	
\usepackage{amssymb}	
\usepackage{float} 		
\usepackage[colorinlistoftodos]{todonotes}
\usepackage{mathtools}
\usepackage{xcolor}
\usepackage{subfig}
\usepackage{hyperref}
\usepackage{tabularx,ragged2e} 
\usepackage{lipsum,afterpage,refcount}
\newcommand{\setfootnotemark}{%
  \refstepcounter{footnote}%
  \footnotemark[\value{footnote}]}
\newcolumntype{C}{>{\Centering\arraybackslash}X}

\newcommand{\msun}{$M_{\odot}$}

\newcommand{\mh}{$M_{\mathrm{H_2}}$}
\newcommand{\mhi}{$M_{\mathrm{HI}}$}
\newcommand{\lco}{$L_\mathrm{{CO(1-0)}}$}
\newcommand{\mstar}{$M_{\ast}$}

\newcommand{\xco}{$\alpha_{\rm CO}$}

\newcommand{\omH}{$\Omega_{\mathrm{H_2}}$}
\newcommand{\omHI}{$\Omega_{\mathrm{HI}}$}
\newcommand{\omHIH}{$\Omega_{\mathrm{HI+H2}}$}
\def\Hmol{$\mathrm{H_2}$}
\def\HI{H\textsc{i}}
\def\HIMF{H\textsc{i}MF}

\defcitealias{lilly2013gas}{L+13}
\defcitealias{mannucci2010fundamental}{M+10}
\defcitealias{bouche2010impact}{B+10}
\defcitealias{dave2012analytic}{Dav\'e+12}
\defcitealias{dekel2013toy}{Dekel+13}
\defcitealias{keres2003co}{K+03}
\defcitealias{obreschkow2009understanding}{O+09}
\defcitealias{popping2015evolution}{P+15}

\title[$\Omega_{\mathrm{H_2}}$ in the local universe]{The cosmic abundance of cold gas in the local Universe}
\author[Fletcher et al.]{
Thomas J. Fletcher$^1$, Am\'elie Saintonge$^1$\thanks{E-mail: a.saintonge@ucl.ac.uk}, Paula S. Soares$^{1, 2}$ and Andrew Pontzen$^1$
\\
$^{1}$ Department of Physics and Astronomy, University College London, Gower Street, London, WC1E 6BT, UK \\
$^{2}$ School of Physics and Astronomy, Queen Mary University of London, Mile End Road, London E1 4NS, UK
}

\date{Accepted XXX. Received YYY; in original form ZZZ}

\pubyear{2020}

\begin{document}
\label{firstpage}
\pagerange{\pageref{firstpage}--\pageref{lastpage}}
\maketitle

\begin{abstract}

We determine the cosmic abundance of molecular hydrogen (\Hmol{}) in the local universe from the xCOLD GASS survey. To constrain the \Hmol{} mass function at low masses and correct for the effect of the lower stellar mass limit of $10^9 \, M_{\odot}$ in the xCOLD GASS survey, we use an empirical approach based on an observed scaling relation between star formation rate and gas mass. We also constrain the \HI{} and \HI{}+\Hmol{} mass functions using the xGASS survey, and compare it to the \HI{} mass function from the ALFALFA survey. We find the cosmic abundance of molecular gas in the local Universe to be $\Omega_{\mathrm{H_2}} = (5.34 \pm 0.47) \times 10^{-5} h^{-1}$. Molecular gas accounts for $19.6\pm3.9\%$ of the total abundance of cold gas,  $\mathrm{\Omega_{HI+H_{2}}} = (4.66 \pm 0.70) \times 10^{-4}\, h_{70}^{-1}$. Galaxies with stellar masses in excess of $10^9$\msun\ account for 89\% of the molecular gas in the local Universe, while in comparison such galaxies only contain 73\% of the cold atomic gas as traced by the \HI{} 21cm line. The xCOLD GASS CO, molecular gas and cold gas mass functions and $\Omega_{\mathrm{H_2}}$ measurements provide constraints for models of galaxy evolution and help to anchor blind ALMA and NOEMA surveys attempting to determine the abundance of molecular gas at high redshifts.

\end{abstract}




\section{Introduction}
\label{intro}

Cold, dense molecular gas is the fuel for star formation in galaxies. The total molecular gas content of a galaxy and its surface density are the main drivers of its star formation and thus evolution \citep{kennicutt1998global}. The interplay between gas inflows and outflows, star formation and galaxy evolution is one of the remaining open issues in astrophysics. Large surveys in the optical, ultraviolet and infrared have clearly established that the overall star formation rate (SFR) of the Universe was significantly higher in the past \citep[e.g.][]{cfrs,galex05,spitzer05,madaureview}. The star formation rate per unit volume was at its highest 8-10 Gyr ago, and has since declined by an order of magnitude to the present day. Extensive work is now going into studying the onset of star formation in the Universe, and tracking the evolution of the cosmic star formation budget during the first 3 Gyr \citep{reddy2009steep, bouwens2012uv, bouwens2015uv, coe2012clash, ellis2012abundance, finkelstein2013galaxy, finkelstein2015evolution, schenker2013uv, oesch2013probing}.

Given that star formation depends on the availability of cold gas, a natural explanation for this observed variation of SFR density with time is that it tracks changes in the amount of cold gas in galaxies that can participate in the star formation process. An alternative explanation would be that the gas contents of galaxies remain roughly unchanged over time, but that the star formation efficiency varies strongly. Observations of molecular gas in galaxies including in local galaxies and those up to $z\sim3$ support the first picture; the cold gas contents of galaxies was significantly higher at the peak of cosmic star formation history \citep{tacconi2013phibss, genzel2015combined} and traces the redshift evolution of the star formation rate, modulo a weak evolution of star formation efficiency \citep{saintonge2013validation}. These results are also reproduced by simple analytical ``equilibrium" models where the SFR is regulated by the mass of the gas reservoir, which is depleted by outflows and replenished by inflowing gas \citep{white1991galaxy, bouche2010impact, dave2011galaxy, dave2012analytic, dekel2013toy, lilly2013gas}.

The implication is that understanding the cosmic star formation history relies on understanding how the cold gas reservoirs of galaxies evolve, both in terms of their total mass, and in the balance between the different phases (cold, warm and hot). In this study, we focus our attention on the cold atomic and molecular gas phases, given their direct link with star formation. In particular, we use the state-of-the art molecular gas survey, xCOLD GASS \citep{saintonge2017}, to accurately calculate the mass function and cosmic abundance of molecular hydrogen.

In the local Universe, the interstellar medium of galaxies is dominated by the cold atomic phase; the atomic-to-molecular mass ratio (\mhi$/$\mh) is on average a factor of $\sim3-4$ in massive galaxies and increases to $\sim10$ in lower mass galaxies (\mstar$=10^9$\msun) \citep{catinella18}. Despite this dominance of the atomic phase, it is nonetheless crucial to accurately determine \omH: (1) Star formation is triggered in cold, dense molecular clouds and therefore \omH\ tells us about the abundance of star-forming gas, unlike \omHI\ which traces large extended reservoirs of atomic gas that are one step removed from the star formation process. For example, galaxies with \HI{} excesses are found to have very low star formation rates \citep{gereb2018multiwavelength}. (2) There are indications that the \omH$/$\omHI\ ratio may be changing with redshift, with \omH{} rising quickly with redshift \citep{decarli2016alma, decarli19, riechers2019coldz} whilst \omHI{} rises more gradually \citep{zafar2013eso, rhee2017neutral}. This picture is also supported by simulations and semi-analytic models (SAMs) \citep{obreschkow2009cosmic, power2010redshift, lagos2011cosmic, popping2015evolution}, making the calibration of the local value critical to anchor other studies.

The cosmic abundance of cold atomic gas, \omHI, and the \HI{} mass function (\HIMF{}) have been measured locally with high accuracy through large blind surveys \citep[e.g.][]{rosenberg02,zwaan2005}, sometimes exploring the specific environments of groups and clusters \citep{kilborn09,kovac09}. The state-of-the art \HI{} mass function has been produced by the ALFALFA survey; with a final catalog of $\sim31500$ HI-detected galaxies with $z<0.06$ \citep{haynes18}, it has the combination of depth and surface area that allows for the accurate determination of both the low-mass and high-mass ends of the \HIMF{} \citep{martin2010arecibo,jones18}.

A large flux-limited blind survey such as ALFALFA is ideal to build a robust and representative gas mass function. The unavailability of a dataset with all these characteristics probing the molecular gas contents of galaxies explains why there have been comparatively fewer measurements of the H$_2$ mass function (H2MF) and \omH. The molecular gas mass of galaxies is most commonly measured from the luminosity of emission lines of the CO molecule, assuming a specific function for the conversion to total molecular gas mass (the so-called CO-to-H$_2$ conversion function, \xco). Consequently, the \Hmol{} mass function is most often derived from targeted surveys for CO.

For example, \cite{keres2003co} (hereafter \citetalias{keres2003co}) used the FCRAO (Five College Radio Astronomy Observatory) Extragalactic CO survey (300 observations including 236 CO detections) \citep{young1995fcrao} to construct a CO luminosity function (COLF) and calculate the value of \omH\ often used as the reference for the local Universe. From the full FCRAO sample, \citetalias{keres2003co} selected galaxies with $S_{60} > 5.24 \, \mathrm{Jy}$
for inclusion in their COLF. This relatively high flux cut-off reduced the sample down to 200 galaxies, with a bias towards infrared-bright and/or nearby galaxies. The consequence of this is an over-representation of starbursting and merging galaxies in the sample, while in fact these rare objects contribute $<15\%$ of the star formation budget of the universe \citep{rodighiero11,sargent12}.

To circumvent such selection biases in the determination of the H2MF, a solution is to use a blind, flux-limited sample. With facilities such as NOEMA, ALMA and the JVLA, relatively small (but deep) fields can be blindly searched for CO emission, making this a viable option only for high redshift studies \citep[e.g.][]{walter2014molecular,pavesi18,decarli19}. Measuring the H2MF through a flux-limited survey is also possible if one chooses to infer the molecular gas mass of galaxies through their far infrared dust emission rather than CO emission lines \citep[e.g.][]{berta13,vallini2016co}.

In this paper, we make use of a volume-limited CO survey, xCOLD GASS \citep{saintonge2017}, to accurately determine the mass function and cosmic abundance of molecular gas at $z\sim0$. There are many key advantages to this approach. First, the sample is stellar-mass selected, and therefore representative of the entire galaxy population with \mstar$>10^9$\msun. With over 500 galaxies, it also has the required statistics to robustly determine the H2MF above the survey's completeness limit. With \Hmol{} masses derived from CO(1-0), it provides an independent measure from the dust-based studies, where inferred gas masses are sensitive to the dust emissivity, the dust-to-gas abundance ratio and contamination along the line of sight.

Throughout this work we assume a cosmology with $H_0 = 70 \mathrm{km\,s^{-1}\, Mpc^{-1}}$, $\Omega_m=0.3$, $\Omega_k=0$ and $ \Omega_{\Lambda}=0.7$.

\section{Data and methods}

\begin{table}
    \centering
    \renewcommand{\arraystretch}{1.1}
	\begin{tabular}{ l c c  }
    \hline
	 \textbf{Catalogue} & \textbf{Low Mass} & \textbf{High Mass}\\
	 \hline
	 Stellar Mass   & $9.0<M_{\ast}<10.0$    & $10.0<M_{\ast}<11.5$\\
	 Redshift       & $0.01<z<0.02$  & $0.025<z<0.05$\\
	 \hline
	 Total galaxies       & 166   & 366\\
	 CO Detections    & 117    & 216\\
	 CO Non-Detections & 49    & 150\\
     \hline
	\end{tabular}
    \caption{Details of the high and low mass xCOLD GASS samples. In total
	532 galaxies were observed for the xCOLD GASS survey.}
    \label{tabledata}
\end{table}

\subsection{Sample and measurements}

xCOLD GASS is a CO(1-0) survey undertaken with the IRAM-30m telescope, targeting 532 mass-selected SDSS galaxies with $M_{*} > 10^9 \,M_{\odot}$. Details about the survey, observations and data products are in \cite{saintonge2011cold1} and \cite{,saintonge2017}, with basic properties of the sample summarised in Table \ref{tabledata}.

The survey meets key requirements for building a mass/luminosity function. First, the xCOLD GASS sample is large and homogeneous, as all the measurements were carried out with the same instrument and a precise observing strategy; all galaxies were observed until the CO(1-0) line was detected, or until a sensitivity to a gas fraction \mh/\mstar\ of $\sim2\%$ was reached \citep{saintonge2011cold1}. This means that even for non-detections, we are able to place stringent upper limits on the molecular gas mass which can be included in our derivation of the H2MF.

Second, the sample is representative of the $z\sim0$ galaxy population with stellar masses larger that $10^9$\msun. The 532 galaxies were selected randomly out of the larger parent sample of all SDSS galaxies in the mass and redshift range of the survey (see Tab. \ref{tabledata}). Therefore unlike in the case of \citetalias{keres2003co}, our sample is not biased towards particularly gas-rich and star-forming galaxies. When selecting the sample, we only required to have a flat distribution in stellar mass, allowing us to study with similar statistics the galaxy population over more than two orders of magnitude in \mstar. This has the advantage of allowing us to constrain the high mass end of the H2MF with better statistics, and we can account for the resulting selection bias using stellar mass function of the SDSS parent sample \citep{saintonge2011cold1}.

\begin{figure}
    \centering
    \includegraphics[width=0.5\textwidth]{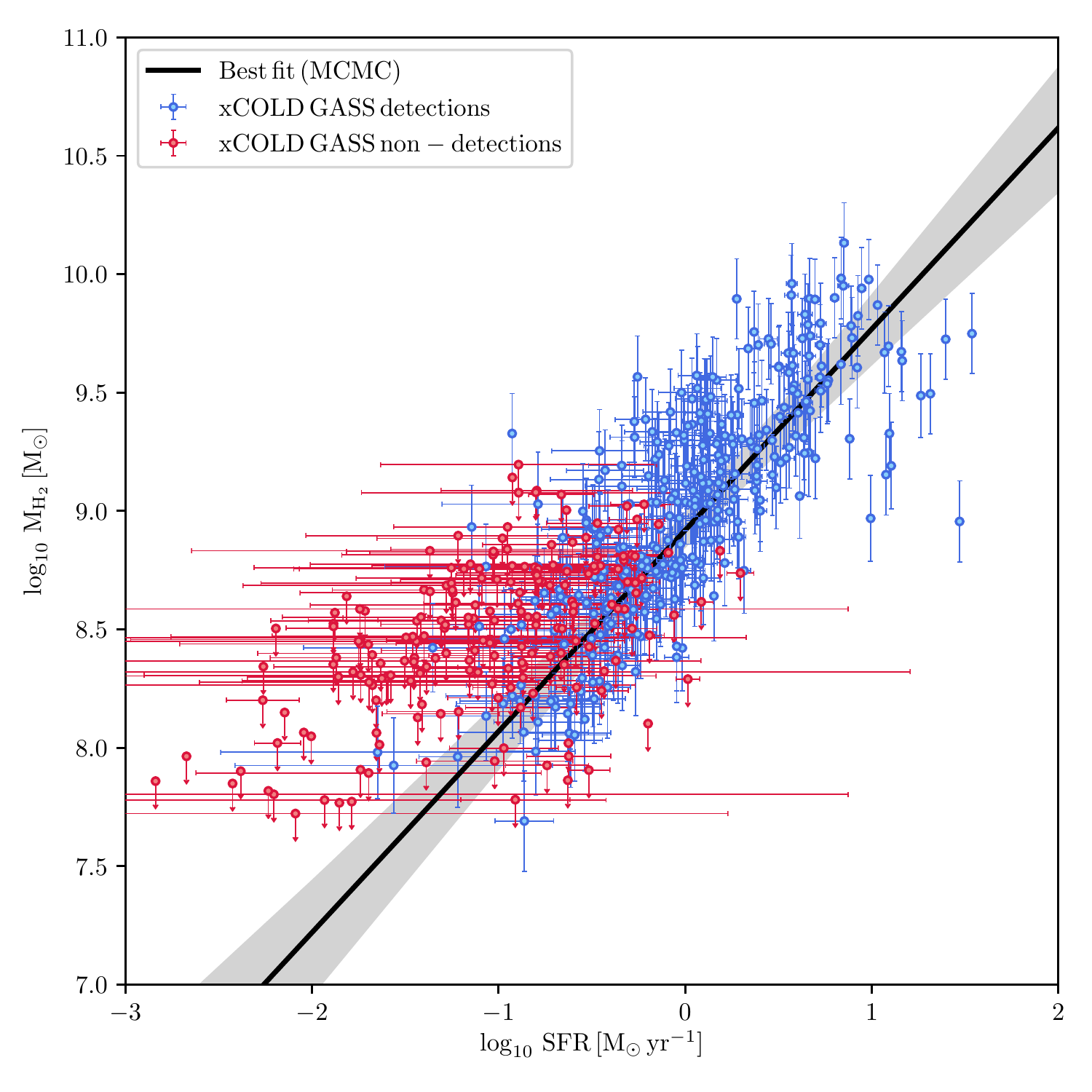}
    \caption{Linear trend between SFR and $M_\mathrm{H_{2}}$ for xCOLD GASS, with detections as blue circles and non-detections as red circles. The black line is the line of best fit, and the shading represents the uncertainty around this fit.}
    \label{fig:sfrH2}
\end{figure}

For each of the 532 xCOLD GASS galaxies we calculate a total molecular gas mass, \mh. For the 333 galaxies where the CO(1-0) line was detected with S/N$>3$, we calculate \mh\ via the CO line luminosity, \lco\ and the CO-to-H$_2$ conversion function of \citet{gio}. The non-detections are treated in three different ways. They are assigned (1) \mh$=0$, which is the most conservative choice, (2) the \mh{} value corresponding to the 5$\sigma$ upper limit, the most optimistic option, and (3) the \mh value expected for the galaxy given its SFR. The third option is motivated by the very well known close relation between molecular gas and star formation, which we parameterise here using the xCOLD GASS sample as shown in Fig. \ref{fig:sfrH2}. We fit the relation between $\log_{10} \, M_\mathrm{H_2}$ and $\log_{10} \, \mathrm{SFR}$ as a first order polynomial, with Gaussian intrinsic scatter around this line with standard deviation $\Lambda$ in the $\log_{10} \, M_\mathrm{H_2}$ direction. This is accomplished using the Markov Chain Monte Carlo (MCMC) affine-invariant ensemble sampler \textit{emcee} \citep{goodman2010, foremanmackey2013}. Unless otherwise stated, throughout this work \textit{emcee} is run using 100 walkers and 1000 steps. The chain for all the parameters is inspected to check for convergence, a burn-in of 500 steps is discarded and the final 100 steps are used. Once the chain has converged, the best-fit relation is:

\begin{equation}
	\log_{10} \,  M_\mathrm{H_{2}} = (0.85 \pm 0.03)\log_{10} \, \mathrm{SFR} + (8.92 \pm 0.02),
	\label{eq:scaling_relation}
\end{equation}

with an intrinsic scatter of $\Lambda = {0.26} \pm {0.02}$. In the case of upper limits, the probability distribution function was integrated from $-\infty$ to the value of the upper limit (see the Appendix of \cite{sawicki2012sedfit} who use the same method). We also use this relation combined with individual metallicity-dependent conversion factors ($\alpha_{\mathrm{CO}}$) to calculate the expected \lco{} for each galaxy for case (3).

\subsection{Building the mass function}
\label{sec:massfunc}

Since xCOLD GASS is not a flux-limited survey, nor a purely volume-limited sample, we must use an appropriate method to build any luminosity or mass function. First, we assign to each xCOLD GASS galaxy a weight to correct for the flat stellar mass distribution of the sample \citep[see Sec.2.2 and Fig.3 in][]{saintonge2017}. Taking these weights into account, the measured values of \mh or \mhi\ are binned and counted to produce a mass ``histogram" that is representative of the entire galaxy population with \mstar$>10^9$\msun.

To then turn this histogram into a mass function (with the correct units of Mpc$^{-3}$~dex$^{-1}$), we compute the effective volume of the survey using the stellar mass function of \citet{baldry2012galaxy}.  The stellar mass function is integrated over the mass range of xCOLD GASS (i.e. \mstar$>10^9$\msun), giving a total inverse volume for galaxies in this range. Finally, to get the total effective volume probed by xCOLD GASS, the total number of galaxies in the survey (532) is divided by this integral. Normalising the mass ``histogram" by this effective volume produces a well calibrated mass function.

\begin{figure*}
    \centering
    \subfloat{{\includegraphics[width=0.45\textwidth]{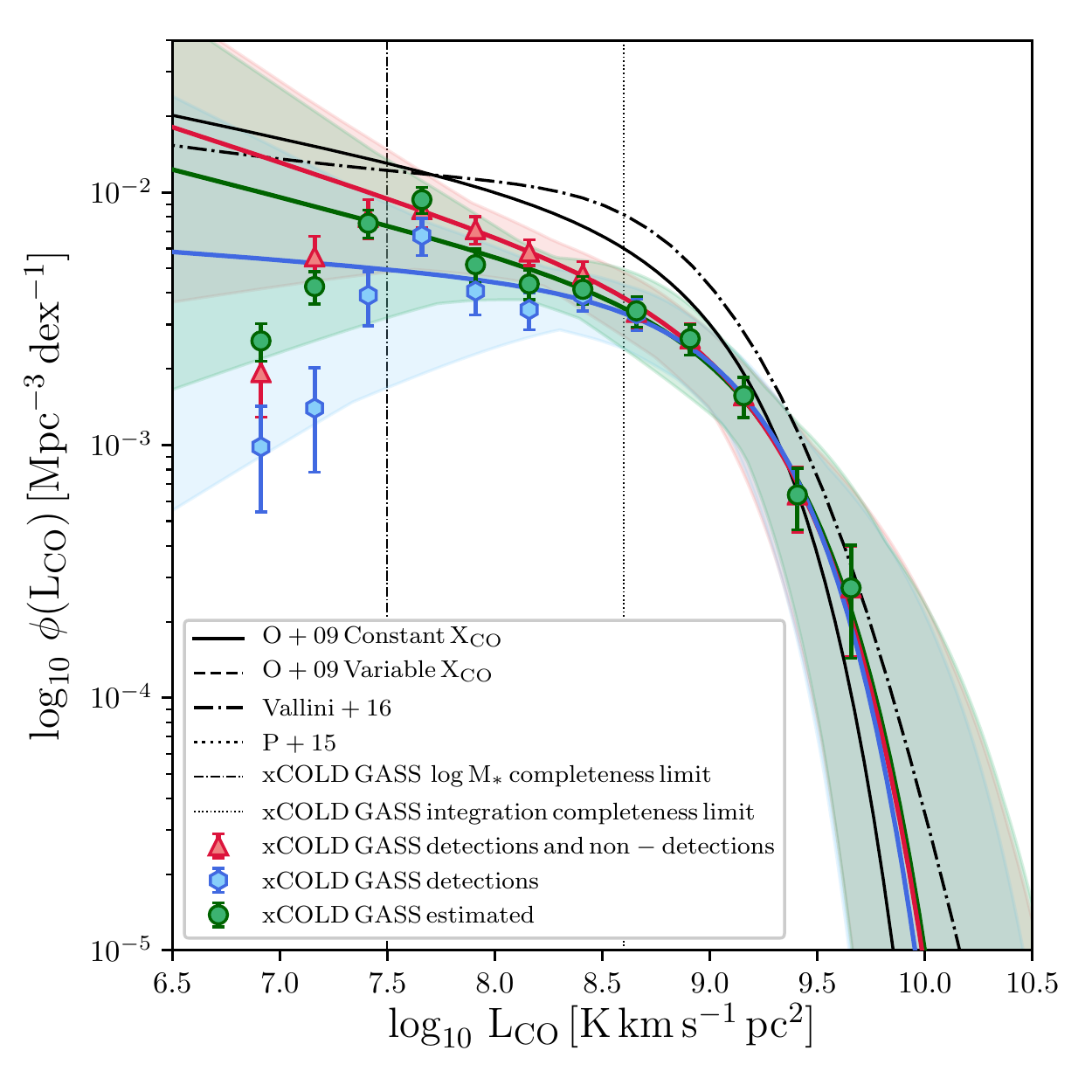}}}%
    \qquad
    \subfloat{{\includegraphics[width=0.45\textwidth]{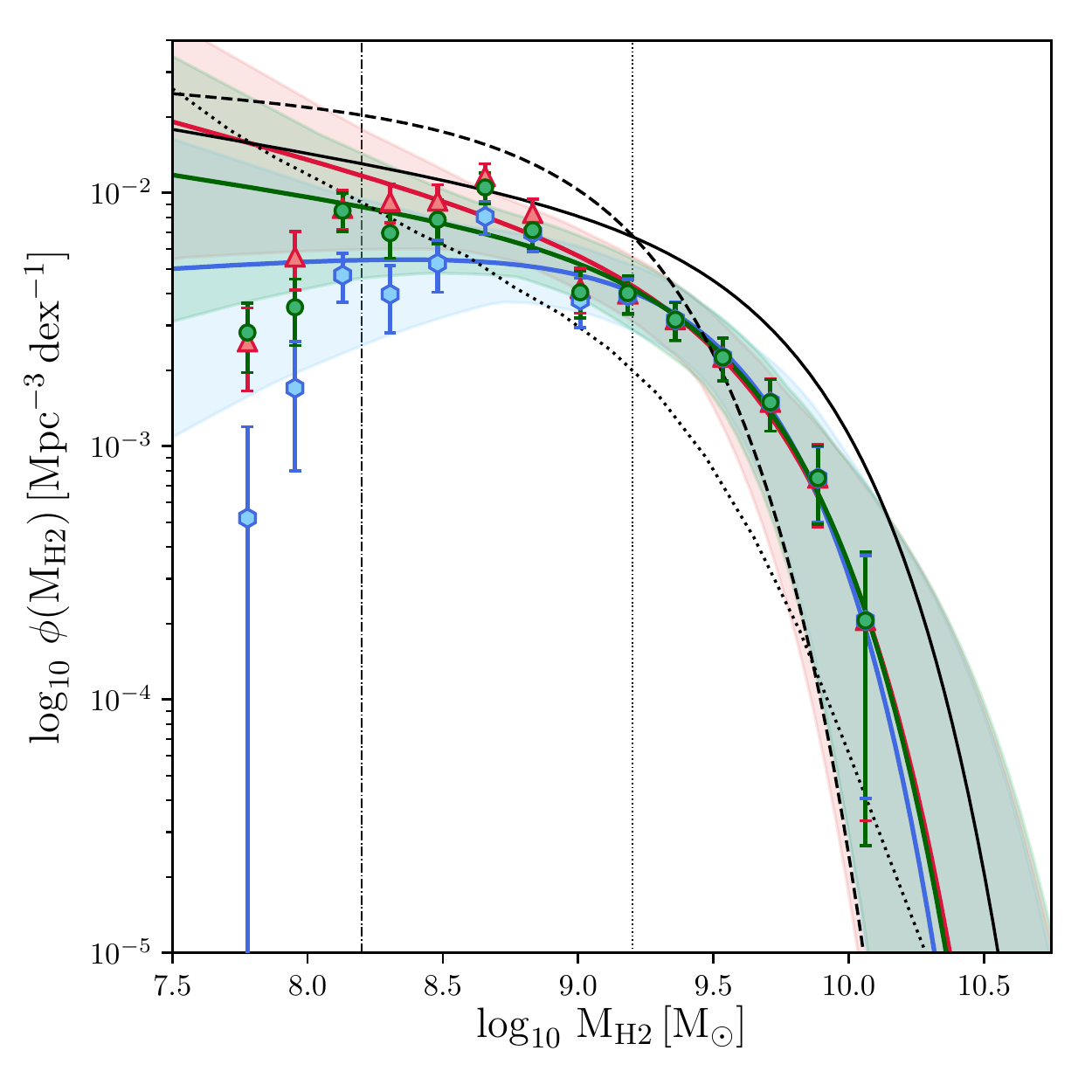}}}%
    \caption{The xCOLD GASS CO luminosity function (COLF, {\it left}) and H$_2$ mass function (H2MF, {\it right}). The CO(1-0) non-detections are treated in three different ways: they are set to zero (blue hexagons), they are set to the 5$\sigma$ upper limit (red triangles), or an estimated value of \lco\ and \mh\ based on the empirical relation shown in Fig. \ref{fig:sfrH2} (green circles). The latter is considered our best and default version. The solid coloured lines and shaded regions show the best fit Schechter function and associated 1$\sigma$ error respectively, for each case. For comparison the \protect\citetalias{obreschkow2009understanding} \Hmol{} Schechter functions using both constant and variable conversion factors are shown (solid black and dashed black lines respectively). A COLF empirically derived from the infrared LF is shown (black dash-dotted line) \citep{vallini2016co}, as well as a \Hmol{} mass function predicted from a semi-analytic model \citetalias{popping2015evolution} (black dotted line). The stellar mass and integration completeness limits are shown as vertical dot-dashed and dotted lines, respectively.}%
    \label{fig:LFMF}%
\end{figure*}

The uncertainty on each point of the luminosity or mass function is calculated using bootstrapping with replacement. For a given sample of $N$ galaxies, we resample $N$ times, allowing for repetitions in the new sample. We also resample each selected galaxy using its measurement error to include the effect of observational errors. Each time we resample the original sample with replacement we also calculate the number density of galaxies in each mass bin and record these values. This procedure is repeated 1000 times and we use the 1$\sigma$ standard deviation of the resamples in each mass bin as the error. Once the mass function is built and errors are determined, the best-fitting Schechter function \citep{Schechter1976} parameters are inferred using \textit{emcee}. The median of the posterior from the converged \textit{emcee} chain is taken as the best fit for each parameter, and the $1\sigma$ error is calculated from the 16th and 84th percentiles. We use the median rather than the mean as the mean can be outside of the $1\sigma$ confidence interval if there is large skewness \citep{hogg2018data}. It should be noted that the median of the posterior probability density function for each parameter, if taken together as a `best-fit' result (\mh{}, $\phi^*$, $\alpha$), does not necessarily result in a best-fit. This is because the median for each parameter is a marginalisation of the posterior probability density function in one dimension. In cases where there is curvature in the parameter space, as is true with Schechter function fits, taking the medians for each parameter collectively can result in a fit that falls in a low-probability region of the posterior probability density function (see \cite{hogg2018data} for a more detailed discussion).

For each luminosity/mass function we build, we determine two completeness limits, which are the consequence of the xCOLD GASS sample selection and observing strategy. First, there is a ``stellar mass completeness" limit, because xCOLD GASS only targeted galaxies with $M_{*} >10^9 M_{\odot}$. Given the positive correlation between \mstar\ and \mh\ (and $L_{\rm CO}$), this completeness limit is set at the highest value of \mh\ found in xCOLD GASS for galaxies near the stellar mass cutoff of $>10^9$\msun. Galaxies with lower stellar masses, which are not present in the xCOLD GASS sample, contribute insignificantly to the luminosity/mass function above that limit. Secondly, the ``integration completeness" limit marks the value of \mh\ (and $L_{\rm CO}$) above which all the galaxies are guaranteed to have a $5\sigma$ detection of the CO(1-0) line, based on the observing strategy of the survey. This is set to \mh$=0.02$\mstar\ for \mstar$=10^{11}$\msun, the mass of the most massive galaxies in the xCOLD GASS sample.

\section{Results}
\label{sec:results}

\subsection{The CO(1-0) luminosity function and molecular gas mass function at $z\sim0$}

The CO luminosity function from xCOLD GASS, computed using the method described above, is presented in {Fig. \ref{fig:LFMF}} (left panel). The three different treatments of the non-detections are used, and the two completeness limits are represented as vertical lines. Above the integration completeness limit, the three versions of the COLF are in agreement, as expected. Below this limit, they begin to clearly diverge. The effect of the stellar mass selection can be clearly seen in the incompleteness below the mass completeness limit. The ``blue" COLF, which has the non-detections all set to \lco{}$=0$, represents a lower limit on the mass function, while the ``red" version is a slight overestimation since it puts all the non-detections at their 5$\sigma$ upper limits. We adopt as our best COLF the ``green" version from Fig. \ref{fig:LFMF}, where we use empirical predictions of \lco\ for the non-detections. The best-fitting\footnote{The converged chains for this and all other inferred Schechter function parameters can be found at {\url{https://github.com/tomjf/Omega_H2_chains}}.} Schechter function above the mass completeness limit is determined with \textit{emcee} and has parameters: $\mathrm{\log_{10}\, {L/K\,km\, s^{-1}\, pc^2}} = 9.29^{+0.14}_{-0.12}$, $\mathrm{\alpha} = -1.25^{+0.10}_{-0.10}$ and $\mathrm{\phi^{*}/10^{-3}Mpc^{-3}} = 1.26^{+0.45}_{-0.38}$.

For comparison, Fig. \ref{fig:LFMF} shows the COLF from the FCRAO survey derived from the H2MF presented in \cite{obreschkow2009understanding} (hereafter \citetalias{obreschkow2009understanding}), as well as a COLF derived from the infrared luminosity function \citep{vallini2016co}. These COLFs only appear to agree for the highest luminosities. The xCOLD GASS result suggests a much lower abundance of intermediate luminosity objects ($7.5<\log_{10}\, L_{\rm CO}<9$) yet a steeper faint end slope compared to the results from \citetalias{obreschkow2009understanding} and \cite{vallini2016co}.

Similarly, the \Hmol{} mass function from xCOLD GASS is shown in Fig. \ref{fig:LFMF} (right-hand panel). If we were using a constant value of $\alpha_{\mathrm{CO}}$ for all galaxies, then the COLF and H2MF would be identical in shape; in particular they would have the same Schechter slope and normalisation parameters $\alpha$ and $\phi$ respectively. Indeed, in Figure \ref{fig:LFMF}, we can see that the mass function using a constant conversion factor from \citetalias{obreschkow2009understanding} has the same shape as the COLF from \citetalias{obreschkow2009understanding}. However, we must take into account that galaxies with sub-solar metallicities are under-luminous in CO, which we accomplish by using the variable $\alpha_{\mathrm{CO}}$ prescription of \citet{gio} to convert from \lco\ to \mh. We also compare our mass function to a H2MF derived using as semi-empirical method  from \cite{popping2015evolution} (hereafter \citetalias{popping2015evolution}) which combines a sub-halo abundance matching model with a model to indirectly estimate the \HI{} and \Hmol{} masses of each galaxy. The $z=0$ mass function data for this model is available for download\footnote{\url{http://www.mpia.de/homes/popping/data/gas_sham.data.tar.gz}}. The \citetalias{popping2015evolution} function does not agree with our results or the \citetalias{obreschkow2009understanding} results around both the knee and the high-mass end. However, this serves to illustrate the range of \Hmol{} mass functions in the literature.

Using this method and again using the ``green'' mass function where non-detections are estimated, the best-fitting Schechter function above the mass completeness limit is determined with \textit{emcee} and has parameters: $\mathrm{log_{10}\, M_{H_2}/M_{\odot}} = 9.59^{+0.11}_{-0.10}$, $\mathrm{\alpha} = -1.18^{+0.11}_{-0.11}$ and $\mathrm{\phi^{*}/10^{-3}Mpc^{-3}} = 2.34^{+0.72}_{-0.61}$.

Comparing our best-fitting Schechter function to the mass function determined using a constant conversion factor from \citetalias{obreschkow2009understanding}, we see that the high-mass end tails off sooner, despite the fact the two bright-ends of the COLFs were in good agreement.  This is because the variable conversion factor prescription decreases with increasing metallicity, giving ($\alpha_{\mathrm{CO}} < 4.36 \, \mathrm{M_{\odot} (K \, km \, s^{-1})^{-1}}$) for galaxies with $\mathrm{12 + \log(O/H)} \gtrsim 8.6$ \citep{gio}, which are typically more massive due to the mass-metallicity relation. Conversely, all galaxies in \citetalias{obreschkow2009understanding} were given a much higher conversion factor of ($\alpha_{\mathrm{CO}} = 6.5 \, \mathrm{M_{\odot}~(K~km~s^{-1})^{-1}}$).

\subsection{The total cold gas (HI + \Hmol{}) mass function}

\begin{figure}
    \centering
    \includegraphics[width=0.5\textwidth]{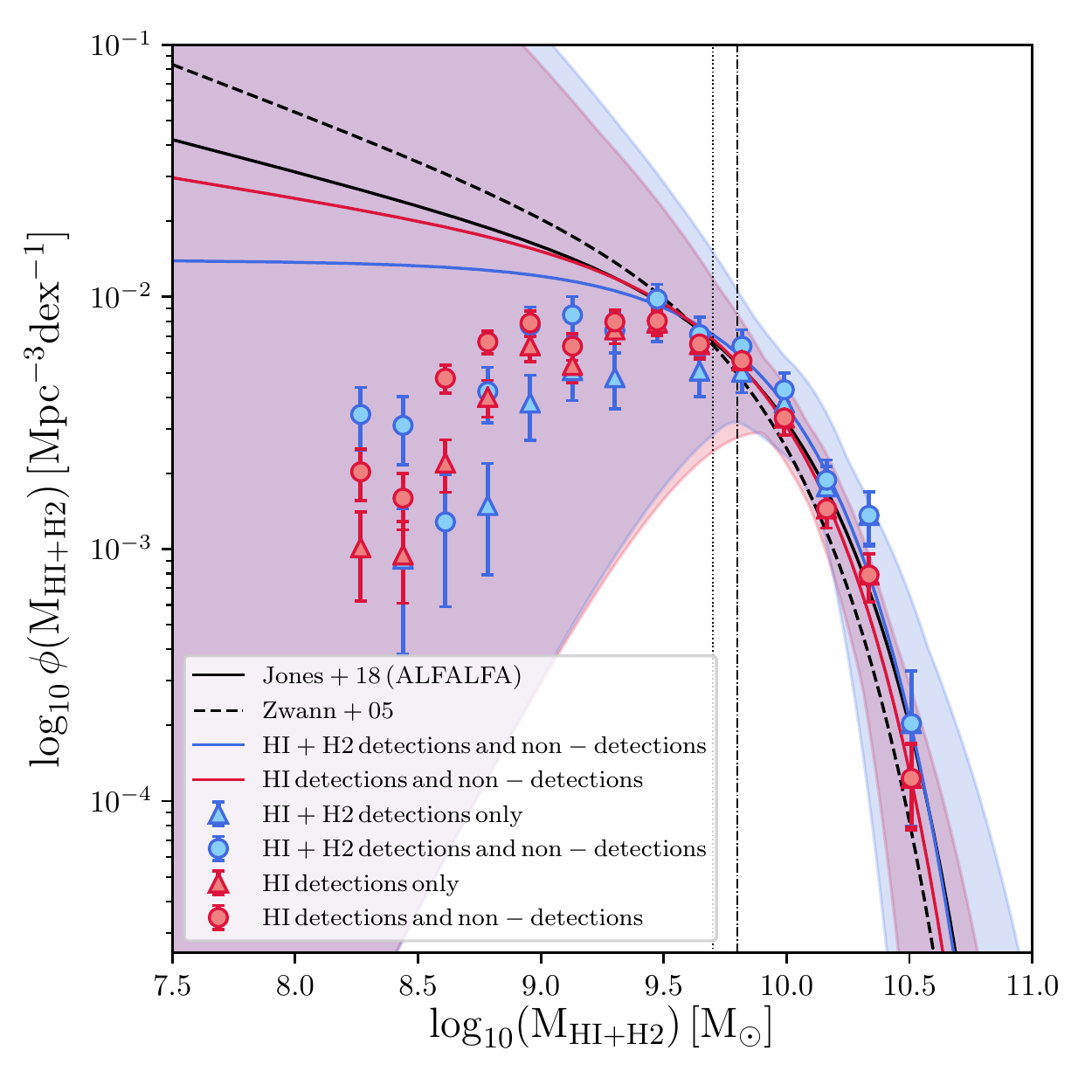}
    \caption{The \HI{} mass function (red) and \HI{} + H$_{2}$ mass function (blue) for the overlapping xGASS and xCOLD GASS sample. All detections and non-detections are shown as filled circles, while detections only are shown as a triangle. The \HI{} mass function derived by \protect\cite{zwaan2005} and the ALFALFA \HIMF{} \protect\citep{jones18} are shown for comparison.}
    \label{fig:HIMF_1}
\end{figure}

Having both \HI{} and \Hmol{} mass measurements for the xCOLD GASS galaxies allows us to determine the mass function of cold gas (\HI{} + \Hmol{}) in the local Universe. Additionally, by building an \HI{} mass function from the xCOLD GASS sample and comparing it to the \HIMF{} from the ALFALFA survey \citep{jones18}, we can validate our methodology (as presented in Section \ref{sec:massfunc}).

The \HI{} data are retrieved from the xGASS catalog \citep{catinella18}, which contains Arecibo \HI{} observations for 1179 SDSS-selected galaxies in the mass range $9.0 < \log_{10}\, M_{*}/\mathrm{M_{\odot}} < 11.5$. Of these galaxies, 477 are also in the xCOLD GASS sample. This is the sub-sample of xCOLD GASS that we use in this section to derive the total \HI{} + \Hmol{} mass function.

In Figure \ref{fig:HIMF_1} we show both the \HIMF{} and total \HI{} + \Hmol{} mass functions derived from xCOLD GASS. In both cases, we give the mass functions for two different treatments of the non-detections: setting them to zero, or assigning them the gas mass equivalent to the 5$\sigma$ upper limit. The completeness limits arising from the stellar mass cutoff and the depth of the observations are shown as before. We fit both mass functions above the stellar mass completion limit with a Schechter function using {\it emcee}. The best-fit parameters are summarized in Table \ref{table:allresults}.

The agreement between the xCOLD GASS and ALFALFA \HIMF{} of \citet{jones18} above our completeness limit is excellent, despite our \HIMF{} being built from a far smaller sample that is mass- rather flux-limited. This further confirms that our methodology for building mass functions from a volume-limited sample (Sec. \ref{sec:massfunc}) is valid, and that our H2MF is accurate in both shape and normalisation. Additionally, it indicates that it is possible to use the ALFALFA \HI{} estimate for $\mathrm{\Omega_{HI}}$ to calculate the total $\mathrm{\Omega_{HI+H_{2}}}$ cold gas abundance in the local universe, since it is compatible with the xCOLD GASS survey above the integration completion limit.

\subsection{The abundance of cold gas at $z \sim 0$}

\begin{figure}
    \centering
    \includegraphics[width=0.5\textwidth]{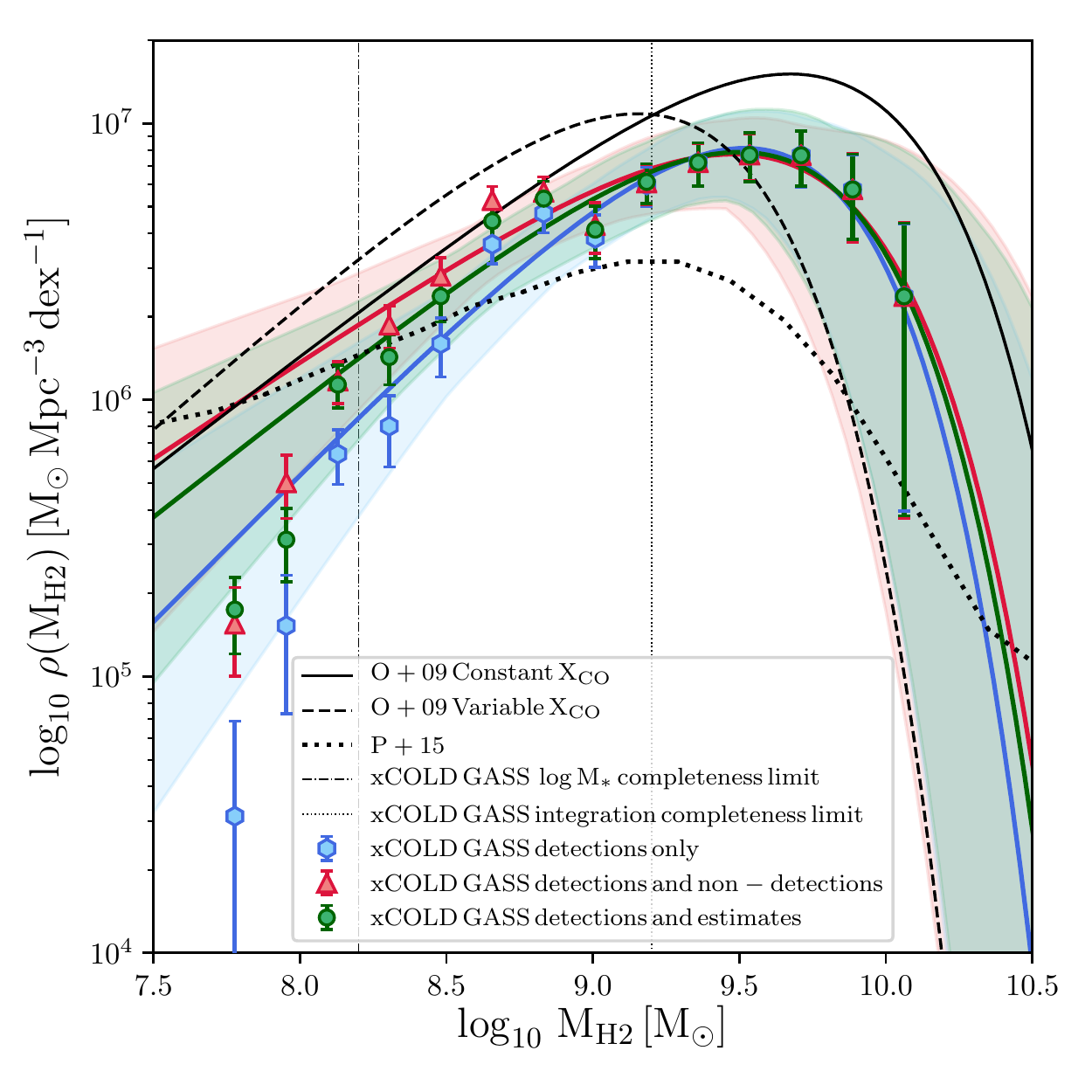}
    \caption{The distribution of $\mathrm{\rho_{H_{2}}}$ with $M_\mathrm{H_{2}}$ for the xCOLD GASS sample. Blue hexagons represent the xCOLD GASS sample of CO(1-0) detections only, red triangles represent xCOLD GASS data with both detections and non-detections, and green circles use the xCOLD GASS detections and estimated masses for the non-detections. The solid coloured lines and shaded regions show the best fit Schechter function and associated 1$\sigma$ error respectively, for each case. The shading around the lines represents the error on the fit. For comparison the \protect\citetalias{obreschkow2009understanding} Schechter functions with constant and variable conversion factors are also shown (solid black and dashed black lines respectively) as well as a result derived from a SAM (\citetalias{popping2015evolution} black dotted line).}
    \label{fig:rhoh2_1}
\end{figure}

The mass functions shown in Fig. \ref{fig:HIMF_1} make it very clear that most of the cold gas in the nearby Universe is in atomic rather than molecular form; we quantify this here by calculating the overall abundance of gas in the different phases. First, in Fig. \ref{fig:rhoh2_1}, we look at the distribution of $\mathrm{\rho_{H_{2}}}$, which is the product of $\phi$(\mh) and \mh. The uncertainties and best fit relations from Fig. \ref{fig:LFMF} are translated over. The total value of $\mathrm{\rho_{H_{2}}}$ (and thus $\mathrm{\Omega_{H_{2}}}$, after dividing by the critical density) can be found by integrating the area underneath the curve in Fig. \ref{fig:rhoh2_1}. For our best treatment of the non-detections, using the empirical scaling relation in Equation \ref{eq:scaling_relation} to predict the true \Hmol{} masses of non-detections, we find a value of $\mathrm{\Omega_{H_{2}}} = (5.34 \pm 0.47)  \times 10^{-5}\, h^{-1}$ (the values for the other two cases are given in Table \ref{table:allresults}). For comparison, summing the bin values derived from xCOLD GASS, which are affected by mass incompleteness, results in a value for \omH{} which is approximately $95\%$ of the value quoted above.

\begin{figure}
    \centering
    \includegraphics[width=0.5\textwidth]{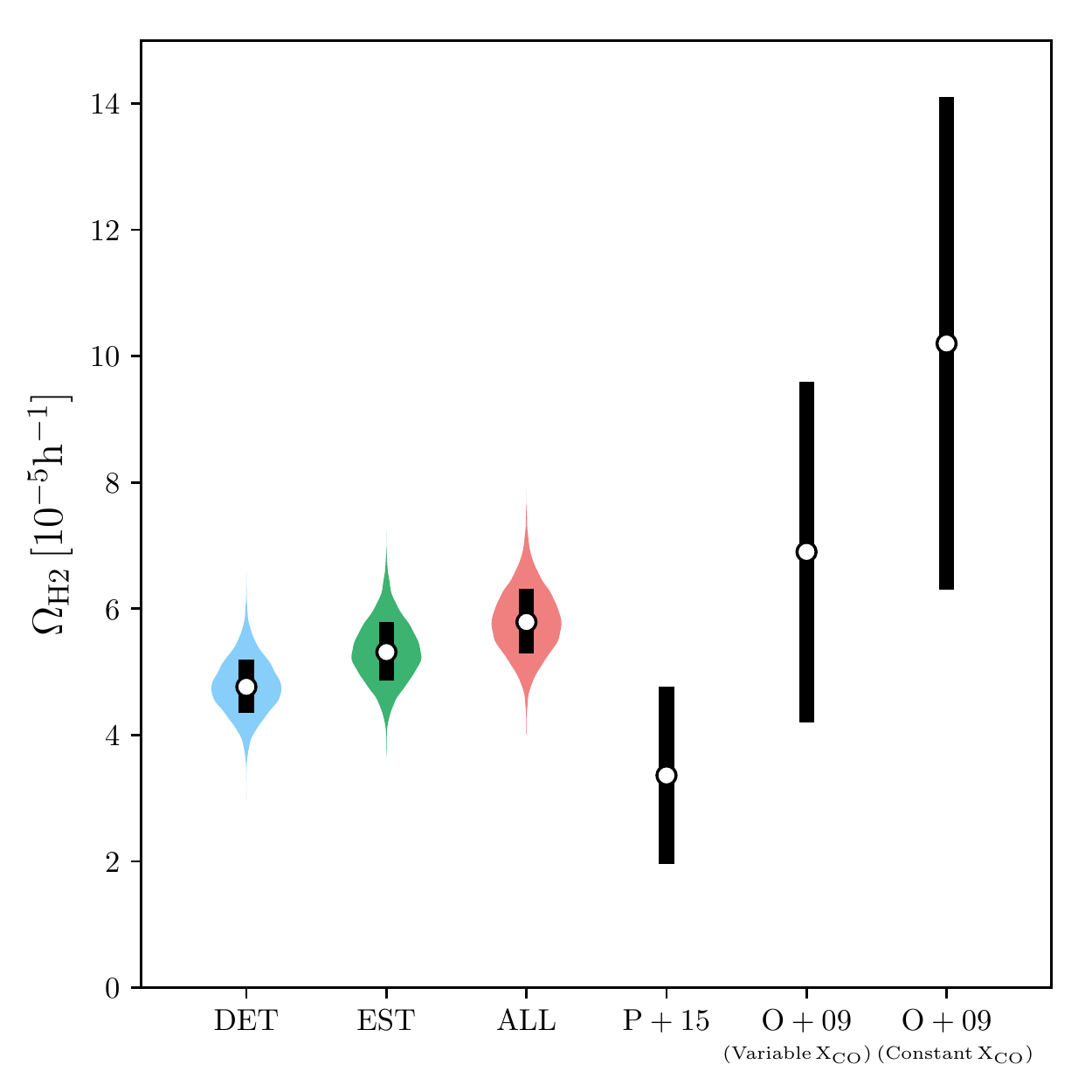}
    \caption{Comparison of \omH{} calculated in this work and previous estimates from the literature. The coloured violin plots show the probability density of \omH{} derived from the the best-fitting Schechter functions. The vertical black lines encompass the 16th to the 84th percentiles, whilst the white dot shows the position of the 50th percentile. The estimates using detections, estimated masses and detections and non-detections are shown as blue, green and red respectively. Three estimates from the literature are shown. Two are derived in \citet{obreschkow2009understanding} using both a constant and variable conversion factor where the third is from \citetalias{popping2015evolution} and is derived using semi-analytic models.}
    \label{fig:omega_H2_summary_plot}
\end{figure}

In Figure \ref{fig:omega_H2_summary_plot}, we compare our new value of $\mathrm{\Omega_{H_{2}}}$ with a range of results for the literature. Our results are within the (large) uncertainties of the early FCRAO work (\citetalias{obreschkow2009understanding}), but lower by a factor of up to 2. This difference comes down to an overestimation in the previous study of the number density of galaxies with \mh$\lesssim10^9$\msun, and different assumptions for the CO-to-H$_2$ conversion factor.

Similarly, we calculate the abundance of \HI{} from our mass function in Fig. \ref{fig:HIMF_1}, and find $\mathrm{\Omega_{HI} = 3.47^{+3.21}_{-0.99} \times 10^{-4}}$. For the ALFALFA survey, \protect\citealt{jones18} find $\mathrm{\Omega_{HI} = (3.9 \pm 0.1 \pm 0.6)\, \times\, 10^{-4}}$ when correcting for \HI{} self-absorption, where the first and second quoted errors are the random and systematic errors respectively. When \HI{} self-absorption is not accounted for $\mathrm{\Omega_{HI-uncorr} = (3.5 \pm 0.1 \pm 0.5)\, \times\, 10^{-4}}$ \citep{jones18}. Our result, albeit with significantly larger uncertainties, is consistent with this value as well as with the HIPASS result \citep{zwaan2005}. Given the higher precision and accuracy of the ALFALFA result, we adopt the \protect\citealt{jones18} value of $\mathrm{\Omega_{HI}}$ and combine with our determination of $\mathrm{\Omega_{H_{2}}}$ to arrive at our best estimate of the total cold gas abundance in the local Universe: $\mathrm{\Omega_{HI+H_{2}}} = (4.66 \pm 0.70) \times 10^{-4}\, h_{70}^{-1}$.

\begin{table*}
\centering
\begin{tabularx}{\linewidth}{l l c c c l}
    \hline
    Schechter Function &
    Knee Units &
    Knee Value &
    Normalization $\phi^{*}/10^{-3}{\rm Mpc}^{-3}$ &
    Slope $\alpha$ &
    Density Parameter \\[1ex]
    \hline
    \\[0.1ex]

    $L_{\mathrm{CO}}$ (DET) &
    $\log_{10} \, \mathrm{L/K \, km \, s^{-1} \, pc^{2}}$ &
    $9.18^{+0.12}_{-0.11}$ &
    $1.73^{+0.51}_{-0.45}$ &
    $-1.07^{+0.11}_{-0.11}$ &
    -- \\[0.8ex]

    $L_{\mathrm{CO}}$ (EST) &
    $\log_{10} \, \mathrm{L/K \, km \, s^{-1} \, pc^{2}}$&
    $9.29^{+0.14}_{-0.12}$ &
    $1.26^{+0.45}_{-0.38}$ &
    $-1.25^{+0.10}_{-0.10}$ &
    --\\[0.8ex]

    $L_{\mathrm{CO}}$ (ALL) &
    $\log_{10} \, \mathrm{L/K \, km \, s^{-1} \, pc^{2}}$ &
    $9.26^{+0.12}_{-0.11}$ &
    $1.37^{+0.45}_{-0.37}$ &
    $-1.27^{+0.09}_{-0.08}$ &
    --\\[0.8ex]

    $M_{\mathrm{H_2}}$ (DET) &
    $\log_{10} \, \mathrm{M_{H_2}/M_{\odot}}$ &
    $9.49^{+0.10}_{-0.09}$ &
    $3.09^{+0.74}_{-0.67}$ &
    $-0.93^{+0.12}_{-0.12}$ &
    $\Omega_{\mathrm{H_2}} = (4.76 \pm 0.43) \times 10^{-5} h^{-1}$\\[0.8ex]

    $M_{\mathrm{H_2}}$ (EST) &
    $\log_{10} \, \mathrm{M_{H_2}/M_{\odot}}$ &
    $9.59^{+0.11}_{-0.10}$ &
    $2.34^{+0.72}_{-0.61}$ &
    $-1.18^{+0.11}_{-0.11}$ &
    $\Omega_{\mathrm{H_2}} = (5.34 \pm 0.47) \times 10^{-5} h^{-1}$ \\[0.8ex]

    $M_{\mathrm{H_2}}$ (ALL) &
    $\log_{10} \, \mathrm{M_{H_2}/M_{\odot}}$ &
    $9.64^{+0.12}_{-0.11}$ &
    $1.99^{+0.69}_{-0.56}$ &
    $-1.30^{+0.10}_{-0.10}$ &
    $\Omega_{\mathrm{H_2}} = (5.82 \pm 0.49) \times 10^{-5} h^{-1}$ \\[0.8ex]

    $M_{\mathrm{HI}}$ &
    $\log_{10} \, \mathrm{M_{HI}/M_{\odot}}$ &
    ${9.86}^{+0.12}_{-0.11}$ &
    ${5.54}^{+1.47}_{-1.77}$ &
    ${-1.16}^{+0.52}_{-0.48}$ &
    $\Omega_{\mathrm{HI}} = 3.47^{+3.21}_{-0.99} \times 10^{-4}$\\[0.8ex]

    $M_{\mathrm{HI+H_{2}}}$ &
    $\log_{10} \, \mathrm{M_{HI+H_2}/M_{\odot}}$ &
    ${9.89}_{-0.16}^{+0.14}$ &
    ${5.97}^{+1.77}_{-2.08}$ &
    ${-1.01}^{+0.68}_{-0.55}$ &
    $\Omega_{\mathrm{HI+H_2}} = 3.64^{+2.99}_{-0.99} \times 10^{-4}$
    \setfootnotemark\label{first}
    \\[0.8ex]
    \hline
\end{tabularx}
\afterpage{\footnotetext[\getrefnumber{first}]{This value for $\Omega_{\mathrm{HI+H_2}}$ is determined using only data above the mass completeness limit for the combined GASS and xCOLD GASS surveys and thus has large uncertainties. Instead we recommend using ${\mathrm{\Omega_{HI+H_{2}}} = (4.66 \pm 0.70) \times 10^{-4}\, h_{70}^{-1}}$ which is the sum of our best estimate for \omH{} and \omHI{} from \cite{jones18}.}}
\caption{All Schechter function and density parameter results.}
\label{table:allresults}
\end{table*}

\subsection{The balance between atomic and molecular gas}

\begin{figure}
    \centering
    \includegraphics[width=0.5\textwidth]{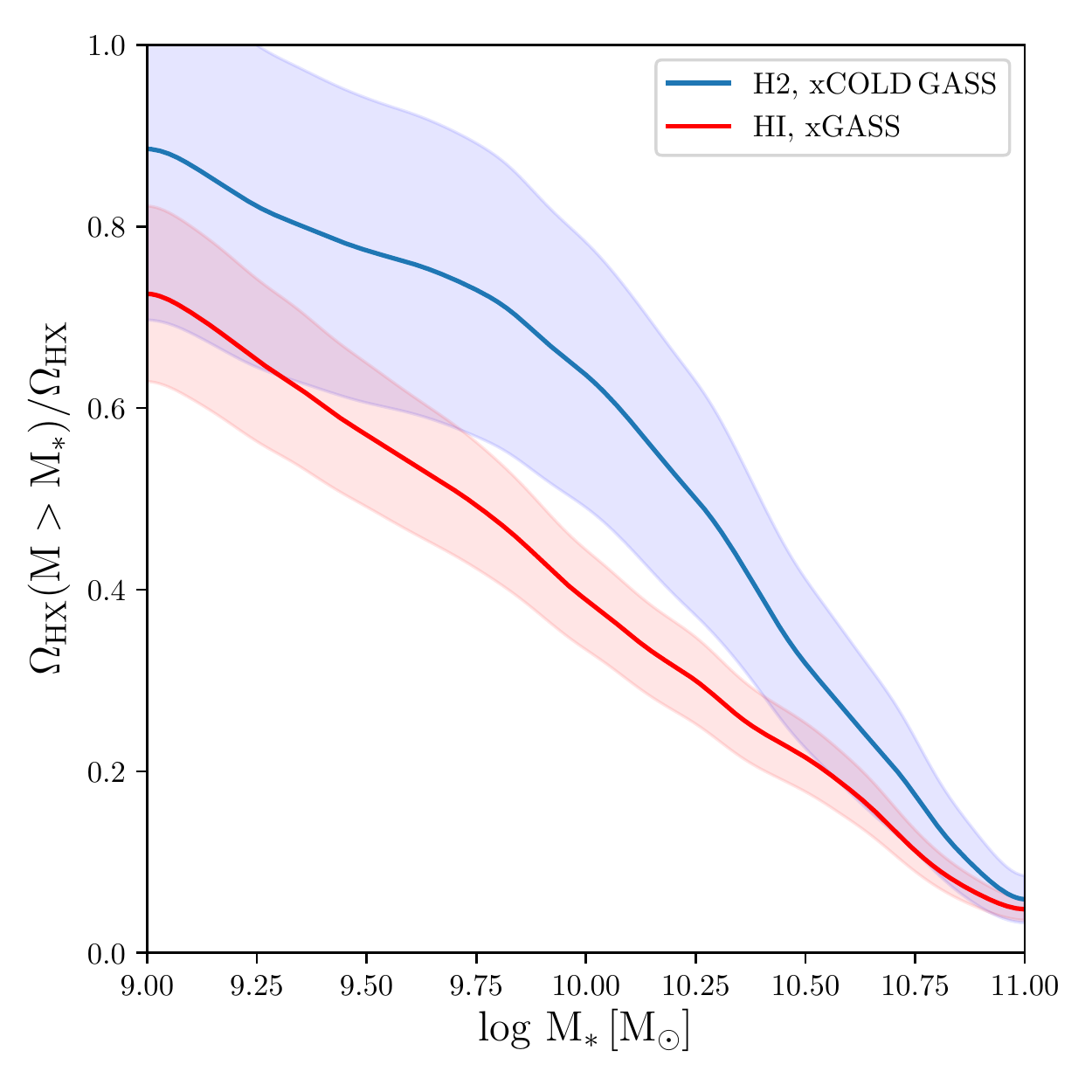}
    \caption{The fraction of $\mathrm{\Omega_{H2}}$ and $\mathrm{\Omega_{HI}}$ coming from galaxies above a given stellar mass. The blue line shows the result for $\mathrm{\Omega_{H2}}$ derived from the xCOLD GASS survey and the red line shows the trend for $\mathrm{\Omega_{HI}}$ using the xGASS survey. The $1\sigma$ error, derived by propagating the bootstrap errors for each H2MF at each step in stellar mass, is shown by the shaded regions.}
    \label{fig:MHI_and_MH2_with_M*}
\end{figure}

The results above show that overall the molecular-to-atomic gas ratio in the local Universe is $19.6 \% \pm 3.9 \%$. An analysis of the xCOLD GASS data has shown how \mh/\mhi\ is a function of stellar mass, with the most massive galaxies ($10^{11}$\msun) having on average a $\sim30\%$ molecular-to-atomic ratio, with the value dropping to $\sim10\%$ for galaxies with stellar masses of $10^{9}$\msun\ \citep{catinella18}. To add to this picture, we show in Figure \ref{fig:MHI_and_MH2_with_M*} the cumulative distribution of the fractional abundance of both \HI{} and H$_2$, normalised by the total cosmic abundance of each gas phase (using our best estimate from xCOLD GASS for $\mathrm{\Omega_{H_{2}}}$ and the ALFALFA value for $\mathrm{\Omega_{HI}}$). The galaxy population with \mstar$>10^9$\msun\ accounts for $\sim 89\%$ of the molecular gas in the local Universe, but only $\sim 73\%$ of the atomic gas as traced by HI. The two lines showing the cumulative distribution of the fractional abundance of both gas phases show different behaviour, with the \HI{} curve growing constantly whilst the \Hmol{} curve is much steeper than the \HI{} curve for \mstar$>10^{10}$\msun\ before becoming shallower than the \HI{} curve at lower masses. This highlights how the ISM of low mass galaxies is atomic gas-dominated. Indeed, while galaxies with \mstar$>10^{10}$\msun\ account for $\sim 64\%$ of all the molecular gas at $z\sim0$, we need to push down to \mstar$\sim10^{9.3}$\msun\ to reach the same level of completeness for the abundance of atomic gas.

The balance between atomic and molecular gas not only varies with stellar mass but may also vary with redshift. Beyond $z=0$ there is growing observational evidence from both CO observations \citep{decarli2016alma, decarli19, riechers2019coldz} and dust-mass tracers \citep{scoville2017evolution} that \omH{} evolves with redshift. Rising to $\sim 6.5$ times the present day value at $z\approx 1-2$ and then either flattening or declining at higher redshifts, mirroring the star-formation history of the universe. Meanwhile observational evidence \citep{zafar2013eso, rhee2017neutral} and simulations \citep{dave2017mufasa} suggest a more gradual evolution of \omHI{} with redshift. Therefore, there is a picture building that at late times \omH{}/\omHI{} declines, which is also supported by SAMs \citep{lagos2011cosmic}. Molecular gas is depleted faster than it is being replenished by \HI{} reservoirs, which could then explain the sudden downturn in SFR since $z\approx 2$. Our determinations for \omH and \omH{}/\omHI{} in this paper can help to provide an anchor at $z=0$ for future studies into the evolution of \omH{} and \omH{}/\omHI{}.

\section{Summary}
\label{sec:discussion}

We present new determinations of the cosmic abundance of molecular gas (\omH{}) and cold gas (\omHIH{}) using the xCOLD GASS survey. The homogeneity, depth of the observations and the stellar-mass selected representative sample from the xCOLD GASS survey allow us to determine the \Hmol{} mass function well below the knee, more accurately than before. In addition we account for the fact that galaxies with sub-solar metallicities are under-luminous in CO by applying a metallicity-dependent conversion factor. Non-detections were treated in three different ways, set to zero, estimated based upon their SFR using a scaling relation between SFR and observed gas mass or set at their $5\sigma$ upper limits. Best-fitting Schechter functions parameters were then inferred using \textit{emcee}, these parameters are then used to calculate the density parameters for \Hmol{} and cold gas. Using this method our best estimate for the cosmic abundance of \Hmol{} is ${\Omega_{\mathrm{H_2}} = (5.34 \pm 0.47) \times 10^{-5} h^{-1}}$. For cold gas we find ${\mathrm{\Omega_{HI+H_{2}}} = (4.66 \pm 0.70) \times 10^{-4}\, h_{70}^{-1}}$. Therefore, we find the ratio of molecular-to-atomic gas as $\Omega_{\rm H2}/\Omega_{\rm HI} = 19.6 \% \pm 3.9 \%$.

Our results provide more stringent constraints on the functional form of the \Hmol{} mass at $z=0$. This is important for cosmological simulations and SAMs which already accurately reproduce the galaxy stellar mass function and its redshift evolution \citep[see][for a review]{somerville2015physical}. However, to do so they must incorporate heuristic models for feedback, which are not well understood and can vary considerably between simulations. As molecular gas studies push to higher redshifts, increasingly, simulations and SAMs must also predict the gas content of galaxies, which is the fuel of star formation thought to drive galaxy evolution \citep{dave2017mufasa, popping2019alma}. We can therefore begin to disseminate between the different heuristic models for feedback, which would not be possible by comparing to the observed galaxy stellar mass function alone. Our best-fitting result: $\mathrm{log M_{H_2}/M_{\odot}} = 9.59^{+0.11}_{-0.10}$, $\mathrm{\alpha} = -1.18^{+0.11}_{-0.11}$ and $\mathrm{\phi^{*}/10^{-3}Mpc^{-3}} = 2.34^{+0.72}_{-0.61}$, for the \Hmol{} mass function and our inferred parameters for the COLF can be used as comparison against predictions made by simulations and SAMs. And this $z=0$ result can be used alongside \omH{} determinations derived using \Hmol{} mass functions at higher redshifts from CO observations \citep{decarli19} or using infrared luminosity as a proxy \citep{vallini2016co}. Furthermore, we have also provided constraints on the total cold gas mass function.

In addition, our results provide determinations of \omH{} and \omH{}/\omHI{} at $z=0$, giving more precise observational anchors for blind ALMA and NOEMA studies investigating the evolution of \Hmol{} at higher redshifts and how this compares to the evolution of \HI{} and SFR. We have also shown the cumulative fractional abundance of atomic and molecular gas, normalised by their respective total cosmic abundances. The two curves of growth for each gas phase show different behaviour, showing that high-mass galaxies tend to be dominated by molecular gas whereas low mass galaxies tend to be dominated by HI.

\section*{Acknowledgements}

We thank Barbara Catinella and Luca Cortese for useful discussion, as well as Dusan Keres and Livia Vallini for advice on the comparison with previous work. TF acknowledges support from STFC. AS and AP acknowledge support from the Royal Society. AP received funding from the European Union's Horizon 2020 research and innovation programme under grant agreement No. 818085 GMGalaxies.



\bsp	
\label{lastpage}

\begin{thebibliography}{}
\makeatletter
\relax
\def\mn@urlcharsother{\let\do\@makeother \do\$\do\&\do\#\do\^\do\_\do\%\do\~}
\def\mn@doi{\begingroup\mn@urlcharsother \@ifnextchar [ {\mn@doi@}
  {\mn@doi@[]}}
\def\mn@doi@[#1]#2{\def\@tempa{#1}\ifx\@tempa\@empty \href
  {http://dx.doi.org/#2} {doi:#2}\else \href {http://dx.doi.org/#2} {#1}\fi
  \endgroup}
\def\mn@eprint#1#2{\mn@eprint@#1:#2::\@nil}
\def\mn@eprint@arXiv#1{\href {http://arxiv.org/abs/#1} {{\tt arXiv:#1}}}
\def\mn@eprint@dblp#1{\href {http://dblp.uni-trier.de/rec/bibtex/#1.xml}
  {dblp:#1}}
\def\mn@eprint@#1:#2:#3:#4\@nil{\def\@tempa {#1}\def\@tempb {#2}\def\@tempc
  {#3}\ifx \@tempc \@empty \let \@tempc \@tempb \let \@tempb \@tempa \fi \ifx
  \@tempb \@empty \def\@tempb {arXiv}\fi \@ifundefined
  {mn@eprint@\@tempb}{\@tempb:\@tempc}{\expandafter \expandafter \csname
  mn@eprint@\@tempb\endcsname \expandafter{\@tempc}}}

\bibitem[\protect\citeauthoryear{{Accurso} et~al.,}{{Accurso}
  et~al.}{2017}]{gio}
{Accurso} G.,  et~al., 2017, \mn@doi [\mnras] {10.1093/mnras/stx1556}, \href
  {https://ui.adsabs.harvard.edu/abs/2017MNRAS.470.4750A} {470, 4750}

\bibitem[\protect\citeauthoryear{Baldry et~al.,}{Baldry
  et~al.}{2012}]{baldry2012galaxy}
Baldry I.,  et~al., 2012, Monthly Notices of the Royal Astronomical Society,
  421, 621

\bibitem[\protect\citeauthoryear{{Berta} et~al.,}{{Berta}
  et~al.}{2013}]{berta13}
{Berta} S.,  et~al., 2013, \mn@doi [\aap] {10.1051/0004-6361/201321776}, \href
  {https://ui.adsabs.harvard.edu/abs/2013A&A...555L...8B} {555, L8}

\bibitem[\protect\citeauthoryear{Bouch{\'e} et~al.,}{Bouch{\'e}
  et~al.}{2010}]{bouche2010impact}
Bouch{\'e} N.,  et~al., 2010, The Astrophysical Journal, 718, 1001

\bibitem[\protect\citeauthoryear{Bouwens et~al.,}{Bouwens
  et~al.}{2012}]{bouwens2012uv}
Bouwens R.,  et~al., 2012, The Astrophysical Journal, 754, 83

\bibitem[\protect\citeauthoryear{Bouwens et~al.,}{Bouwens
  et~al.}{2015}]{bouwens2015uv}
Bouwens R.,  et~al., 2015, The Astrophysical Journal, 803, 34

\bibitem[\protect\citeauthoryear{{Catinella} et~al.,}{{Catinella}
  et~al.}{2018}]{catinella18}
{Catinella} B.,  et~al., 2018, Monthly Notices of the Royal Astronomical
  Society, 476, 875

\bibitem[\protect\citeauthoryear{Coe et~al.,}{Coe et~al.}{2012}]{coe2012clash}
Coe D.,  et~al., 2012, The Astrophysical Journal, 762, 32

\bibitem[\protect\citeauthoryear{Dav{\'e}, Finlator  \& Oppenheimer}{Dav{\'e}
  et~al.}{2011}]{dave2011galaxy}
Dav{\'e} R.,  Finlator K.,   Oppenheimer B.~D.,  2011, Monthly Notices of the
  Royal Astronomical Society, 416, 1354

\bibitem[\protect\citeauthoryear{Dav{\'e}, Finlator  \& Oppenheimer}{Dav{\'e}
  et~al.}{2012}]{dave2012analytic}
Dav{\'e} R.,  Finlator K.,   Oppenheimer B.~D.,  2012, Monthly Notices of the
  Royal Astronomical Society, 421, 98

\bibitem[\protect\citeauthoryear{Dav{\'e}, Rafieferantsoa, Thompson  \&
  Hopkins}{Dav{\'e} et~al.}{2017}]{dave2017mufasa}
Dav{\'e} R.,  Rafieferantsoa M.~H.,  Thompson R.~J.,   Hopkins P.~F.,  2017,
  Monthly Notices of the Royal Astronomical Society, 467, 115

\bibitem[\protect\citeauthoryear{{Decarli} et~al.,}{{Decarli}
  et~al.}{2016}]{decarli2016alma}
{Decarli} R.,  et~al., 2016, \mn@doi [\apj] {10.3847/1538-4357/833/1/69}, \href
  {https://ui.adsabs.harvard.edu/abs/2016ApJ...833...69D} {833, 69}

\bibitem[\protect\citeauthoryear{{Decarli} et~al.,}{{Decarli}
  et~al.}{2019}]{decarli19}
{Decarli} R.,  et~al., 2019, \mn@doi [\apj] {10.3847/1538-4357/ab30fe}, \href
  {https://ui.adsabs.harvard.edu/abs/2019ApJ...882..138D} {882, 138}

\bibitem[\protect\citeauthoryear{Dekel, Zolotov, Tweed, Cacciato, Ceverino  \&
  Primack}{Dekel et~al.}{2013}]{dekel2013toy}
Dekel A.,  Zolotov A.,  Tweed D.,  Cacciato M.,  Ceverino D.,   Primack J.,
  2013, Monthly Notices of the Royal Astronomical Society, 435, 999

\bibitem[\protect\citeauthoryear{Ellis et~al.,}{Ellis
  et~al.}{2012}]{ellis2012abundance}
Ellis R.~S.,  et~al., 2012, The Astrophysical Journal Letters, 763, L7

\bibitem[\protect\citeauthoryear{Finkelstein et~al.,}{Finkelstein
  et~al.}{2013}]{finkelstein2013galaxy}
Finkelstein S.~L.,  et~al., 2013, Nature, 502, 524

\bibitem[\protect\citeauthoryear{Finkelstein et~al.,}{Finkelstein
  et~al.}{2015}]{finkelstein2015evolution}
Finkelstein S.~L.,  et~al., 2015, The Astrophysical Journal, 810, 71

\bibitem[\protect\citeauthoryear{{Foreman-Mackey}, {Hogg}, {Lang}  \&
  {Goodman}}{{Foreman-Mackey} et~al.}{2013}]{foremanmackey2013}
{Foreman-Mackey} D.,  {Hogg} D.~W.,  {Lang} D.,   {Goodman} J.,  2013,
  Publications of the Astronomical Society of the Pacific, 125, 306

\bibitem[\protect\citeauthoryear{Genzel et~al.,}{Genzel
  et~al.}{2015}]{genzel2015combined}
Genzel R.,  et~al., 2015, The Astrophysical Journal, 800, 20

\bibitem[\protect\citeauthoryear{Ger{\'e}b, Janowiecki, Catinella, Cortese  \&
  Kilborn}{Ger{\'e}b et~al.}{2018}]{gereb2018multiwavelength}
Ger{\'e}b K.,  Janowiecki S.,  Catinella B.,  Cortese L.,   Kilborn V.,  2018,
  Monthly Notices of the Royal Astronomical Society, 476, 896

\bibitem[\protect\citeauthoryear{{Goodman} \& {Weare}}{{Goodman} \&
  {Weare}}{2010}]{goodman2010}
{Goodman} J.,  {Weare} J.,  2010, Communications in Applied Mathematics and
  Computational Science, 5, 65

\bibitem[\protect\citeauthoryear{{Haynes} et~al.,}{{Haynes}
  et~al.}{2018}]{haynes18}
{Haynes} M.~P.,  et~al., 2018, \mn@doi [\apj] {10.3847/1538-4357/aac956}, \href
  {https://ui.adsabs.harvard.edu/abs/2018ApJ...861...49H} {861, 49}

\bibitem[\protect\citeauthoryear{Hogg \& Foreman-Mackey}{Hogg \&
  Foreman-Mackey}{2018}]{hogg2018data}
Hogg D.~W.,  Foreman-Mackey D.,  2018, The Astrophysical Journal Supplement
  Series, 236, 11

\bibitem[\protect\citeauthoryear{{Jones}, {Haynes}, {Giovanelli}  \&
  {Moorman}}{{Jones} et~al.}{2018}]{jones18}
{Jones} M.~G.,  {Haynes} M.~P.,  {Giovanelli} R.,   {Moorman} C.,  2018,
  \mn@doi [\mnras] {10.1093/mnras/sty521}, \href
  {https://ui.adsabs.harvard.edu/abs/2018MNRAS.477....2J} {477, 2}

\bibitem[\protect\citeauthoryear{Kennicutt~Jr}{Kennicutt~Jr}{1998}]{kennicutt1998global}
Kennicutt~Jr R.~C.,  1998, The Astrophysical Journal, 498, 541

\bibitem[\protect\citeauthoryear{Keres, Yun  \& Young}{Keres
  et~al.}{2003}]{keres2003co}
Keres D.,  Yun M.~S.,   Young J.,  2003, The Astrophysical Journal, 582, 659

\bibitem[\protect\citeauthoryear{{Kilborn}, {Forbes}, {Barnes}, {Koribalski},
  {Brough}  \& {Kern}}{{Kilborn} et~al.}{2009}]{kilborn09}
{Kilborn} V.~A.,  {Forbes} D.~A.,  {Barnes} D.~G.,  {Koribalski} B.~S.,
  {Brough} S.,   {Kern} K.,  2009, \mn@doi [\mnras]
  {10.1111/j.1365-2966.2009.15587.x}, \href
  {https://ui.adsabs.harvard.edu/abs/2009MNRAS.400.1962K} {400, 1962}

\bibitem[\protect\citeauthoryear{{Kova{\v{c}}}, {Oosterloo}  \& {van der
  Hulst}}{{Kova{\v{c}}} et~al.}{2009}]{kovac09}
{Kova{\v{c}}} K.,  {Oosterloo} T.~A.,   {van der Hulst} J.~M.,  2009, \mn@doi
  [\mnras] {10.1111/j.1365-2966.2009.14662.x}, \href
  {https://ui.adsabs.harvard.edu/abs/2009MNRAS.400..743K} {400, 743}

\bibitem[\protect\citeauthoryear{Lagos, Baugh, Lacey, Benson, Kim  \&
  Power}{Lagos et~al.}{2011}]{lagos2011cosmic}
Lagos C. d.~P.,  Baugh C.~M.,  Lacey C.~G.,  Benson A.~J.,  Kim H.-S.,   Power
  C.,  2011, Monthly Notices of the Royal Astronomical Society, 418, 1649

\bibitem[\protect\citeauthoryear{{Le Floc'h} et~al.,}{{Le Floc'h}
  et~al.}{2005}]{spitzer05}
{Le Floc'h} E.,  et~al., 2005, \mn@doi [\apj] {10.1086/432789}, \href
  {https://ui.adsabs.harvard.edu/abs/2005ApJ...632..169L} {632, 169}

\bibitem[\protect\citeauthoryear{{Lilly}, {Le Fevre}, {Hammer}  \&
  {Crampton}}{{Lilly} et~al.}{1996}]{cfrs}
{Lilly} S.~J.,  {Le Fevre} O.,  {Hammer} F.,   {Crampton} D.,  1996, ApJL, 460,
  L1

\bibitem[\protect\citeauthoryear{Lilly, Carollo, Pipino, Renzini  \&
  Peng}{Lilly et~al.}{2013}]{lilly2013gas}
Lilly S.~J.,  Carollo C.~M.,  Pipino A.,  Renzini A.,   Peng Y.,  2013, The
  Astrophysical Journal, 772, 119

\bibitem[\protect\citeauthoryear{{Madau} \& {Dickinson}}{{Madau} \&
  {Dickinson}}{2014}]{madaureview}
{Madau} P.,  {Dickinson} M.,  2014, \mn@doi [\araa]
  {10.1146/annurev-astro-081811-125615}, \href
  {https://ui.adsabs.harvard.edu/abs/2014ARA&A..52..415M} {52, 415}

\bibitem[\protect\citeauthoryear{Martin, Papastergis, Giovanelli, Haynes,
  Springob  \& Stierwalt}{Martin et~al.}{2010}]{martin2010arecibo}
Martin A.~M.,  Papastergis E.,  Giovanelli R.,  Haynes M.~P.,  Springob C.~M.,
   Stierwalt S.,  2010, The Astrophysical Journal, 723, 1359

\bibitem[\protect\citeauthoryear{Obreschkow \& Rawlings}{Obreschkow \&
  Rawlings}{2009a}]{obreschkow2009understanding}
Obreschkow D.,  Rawlings S.,  2009a, Monthly Notices of the Royal Astronomical
  Society, 394, 1857

\bibitem[\protect\citeauthoryear{Obreschkow \& Rawlings}{Obreschkow \&
  Rawlings}{2009b}]{obreschkow2009cosmic}
Obreschkow D.,  Rawlings S.,  2009b, The Astrophysical Journal Letters, 696,
  L129

\bibitem[\protect\citeauthoryear{Oesch et~al.,}{Oesch
  et~al.}{2013}]{oesch2013probing}
Oesch P.,  et~al., 2013, The Astrophysical Journal, 773, 75

\bibitem[\protect\citeauthoryear{{Pavesi} et~al.,}{{Pavesi}
  et~al.}{2018}]{pavesi18}
{Pavesi} R.,  et~al., 2018, \mn@doi [\apj] {10.3847/1538-4357/aacb79}, \href
  {https://ui.adsabs.harvard.edu/abs/2018ApJ...864...49P} {864, 49}

\bibitem[\protect\citeauthoryear{Popping, Behroozi  \& Peeples}{Popping
  et~al.}{2015}]{popping2015evolution}
Popping G.,  Behroozi P.~S.,   Peeples M.~S.,  2015, Monthly Notices of the
  Royal Astronomical Society, 449, 477

\bibitem[\protect\citeauthoryear{Popping et~al.,}{Popping
  et~al.}{2019}]{popping2019alma}
Popping G.,  et~al., 2019, arXiv preprint arXiv:1903.09158

\bibitem[\protect\citeauthoryear{Power, Baugh  \& Lacey}{Power
  et~al.}{2010}]{power2010redshift}
Power C.,  Baugh C.,   Lacey C.,  2010, Monthly Notices of the Royal
  Astronomical Society, 406, 43

\bibitem[\protect\citeauthoryear{Reddy \& Steidel}{Reddy \&
  Steidel}{2009}]{reddy2009steep}
Reddy N.~A.,  Steidel C.~C.,  2009, The Astrophysical Journal, 692, 778

\bibitem[\protect\citeauthoryear{Rhee, Lah, Briggs, Chengalur, Colless,
  Willner, Ashby  \& Le~F{\`e}vre}{Rhee et~al.}{2017}]{rhee2017neutral}
Rhee J.,  Lah P.,  Briggs F.~H.,  Chengalur J.~N.,  Colless M.,  Willner S.~P.,
   Ashby M.~L.,   Le~F{\`e}vre O.,  2017, Monthly Notices of the Royal
  Astronomical Society, 473, 1879

\bibitem[\protect\citeauthoryear{Riechers et~al.,}{Riechers
  et~al.}{2019}]{riechers2019coldz}
Riechers D.~A.,  et~al., 2019, The Astrophysical Journal, 872, 7

\bibitem[\protect\citeauthoryear{{Rodighiero} et~al.,}{{Rodighiero}
  et~al.}{2011}]{rodighiero11}
{Rodighiero} G.,  et~al., 2011, \mn@doi [\apjl] {10.1088/2041-8205/739/2/L40},
  \href {http://adsabs.harvard.edu/abs/2011ApJ...739L..40R} {739, L40+}

\bibitem[\protect\citeauthoryear{{Rosenberg} \& {Schneider}}{{Rosenberg} \&
  {Schneider}}{2002}]{rosenberg02}
{Rosenberg} J.~L.,  {Schneider} S.~E.,  2002, \mn@doi [\apj] {10.1086/338377},
  \href {https://ui.adsabs.harvard.edu/abs/2002ApJ...567..247R} {567, 247}

\bibitem[\protect\citeauthoryear{Saintonge et~al.,}{Saintonge
  et~al.}{2011}]{saintonge2011cold1}
Saintonge A.,  et~al., 2011, Monthly Notices of the Royal Astronomical Society,
  415, 32

\bibitem[\protect\citeauthoryear{Saintonge et~al.,}{Saintonge
  et~al.}{2013}]{saintonge2013validation}
Saintonge A.,  et~al., 2013, The Astrophysical Journal, 778, 2

\bibitem[\protect\citeauthoryear{{Saintonge} et~al.,}{{Saintonge}
  et~al.}{2017}]{saintonge2017}
{Saintonge} A.,  et~al., 2017, The Astrophysical Journal Supplement, 233, 22

\bibitem[\protect\citeauthoryear{{Sargent}, {B{\'e}thermin}, {Daddi}  \&
  {Elbaz}}{{Sargent} et~al.}{2012}]{sargent12}
{Sargent} M.~T.,  {B{\'e}thermin} M.,  {Daddi} E.,   {Elbaz} D.,  2012, \mn@doi
  [\apjl] {10.1088/2041-8205/747/2/L31}, \href
  {http://adsabs.harvard.edu/abs/2012ApJ...747L..31S} {747, L31}

\bibitem[\protect\citeauthoryear{Sawicki}{Sawicki}{2012}]{sawicki2012sedfit}
Sawicki M.,  2012, Publications of the Astronomical Society of the Pacific,
  124, 1208

\bibitem[\protect\citeauthoryear{{Schechter}}{{Schechter}}{1976}]{Schechter1976}
{Schechter} P.,  1976, The Astrophysical Journal, 203, 297

\bibitem[\protect\citeauthoryear{Schenker et~al.,}{Schenker
  et~al.}{2013}]{schenker2013uv}
Schenker M.~A.,  et~al., 2013, The Astrophysical Journal, 768, 196

\bibitem[\protect\citeauthoryear{{Schiminovich} et~al.,}{{Schiminovich}
  et~al.}{2005}]{galex05}
{Schiminovich} D.,  et~al., 2005, \apjl, 619, L47

\bibitem[\protect\citeauthoryear{Scoville et~al.,}{Scoville
  et~al.}{2017}]{scoville2017evolution}
Scoville N.,  et~al., 2017, The Astrophysical Journal, 837, 150

\bibitem[\protect\citeauthoryear{Somerville \& Dav{\'e}}{Somerville \&
  Dav{\'e}}{2015}]{somerville2015physical}
Somerville R.~S.,  Dav{\'e} R.,  2015, Annual Review of Astronomy and
  Astrophysics, 53, 51

\bibitem[\protect\citeauthoryear{Tacconi et~al.,}{Tacconi
  et~al.}{2013}]{tacconi2013phibss}
Tacconi L.,  et~al., 2013, The Astrophysical Journal, 768, 74

\bibitem[\protect\citeauthoryear{Vallini, Gruppioni, Pozzi, Vignali  \&
  Zamorani}{Vallini et~al.}{2016}]{vallini2016co}
Vallini L.,  Gruppioni C.,  Pozzi F.,  Vignali C.,   Zamorani G.,  2016,
  Monthly Notices of the Royal Astronomical Society: Letters, 456, L40

\bibitem[\protect\citeauthoryear{Walter et~al.,}{Walter
  et~al.}{2014}]{walter2014molecular}
Walter F.,  et~al., 2014, The Astrophysical Journal, 782, 79

\bibitem[\protect\citeauthoryear{White \& Frenk}{White \&
  Frenk}{1991}]{white1991galaxy}
White S.~D.,  Frenk C.~S.,  1991, The Astrophysical Journal, 379, 52

\bibitem[\protect\citeauthoryear{Young et~al.,}{Young
  et~al.}{1995}]{young1995fcrao}
Young J.~S.,  et~al., 1995, The Astrophysical Journal Supplement Series, 98,
  219

\bibitem[\protect\citeauthoryear{Zafar, P{\'e}roux, Popping, Milliard,
  Deharveng  \& Frank}{Zafar et~al.}{2013}]{zafar2013eso}
Zafar T.,  P{\'e}roux C.,  Popping A.,  Milliard B.,  Deharveng J.-M.,   Frank
  S.,  2013, Astronomy \& Astrophysics, 556, A141

\bibitem[\protect\citeauthoryear{{Zwaan}, {Meyer}, {Staveley-Smith}  \&
  {Webster}}{{Zwaan} et~al.}{2005}]{zwaan2005}
{Zwaan} M.~A.,  {Meyer} M.~J.,  {Staveley-Smith} L.,   {Webster} R.~L.,  2005,
  Monthly Notices of the Royal Astronomical Society, 359, 30

\makeatother
\end{thebibliography}
\end{document}